\documentstyle[12pt]{article}
\newcommand{\noi}{\noindent}
\newcommand{\eq}{\begin{equation}}
\newcommand{\en}{\end{equation}}
\newcommand{\eqa}{\begin{eqnarray}}
\newcommand{\ena}{\end{eqnarray}}

\newcommand{\ra}{\rightarrow}

\begin{document}
\hbox{}
\noindent November 30, 1994 \hfill HU Berlin--IEP--94/19

\hbox{}
                              \hfill Bielefeld BI-TP--94/47

\begin{center}
\vspace*{1.5cm}
\renewcommand{\thefootnote}{\fnsymbol{footnote}}
\setcounter{footnote}{1}
{\LARGE Polyakov loop and spin correlators on finite lattices}
%--------------------------------------------------------------------
\footnote{Work supported by the Deutsche
Forschungsgemeinschaft under research grants Mu 932/1-3 and
Pe 340/3-2. Partial support by EC contract CHRX-CT92-0051
is acknowledged.} \\
\vspace*{0.5cm}
{\large A study beyond the mass gap}

\vspace*{1.5cm}
{\large
J.~Engels $\mbox{}^1$,
V. K.~Mitrjushkin $\mbox{}^2$ \footnote{Permanent address:
Joint Institute for Nuclear Research, Dubna, Russia} and
T.~Neuhaus $\mbox{}^1$
}\\
\vspace*{0.7cm}
{\normalsize
$\mbox{}^1$ {\em Fakult\"{a}t f\"{u}r Physik, Universit\"{a}t
Bielefeld, 33615 Bielefeld, Germany}\\
$\mbox{}^2$ {\em Fachbereich Physik, Humboldt-Universit\"{a}t,
10099 Berlin, Germany}\\     %end normalsize
}
\end{center}
%==============================================================================
\vspace{0.5cm}
\begin{abstract}
 We derive an analytic expression for point-to-point correlation
functions of the Polyakov loop based on the transfer matrix formalism.
For the $2d$ Ising model we show that the results deduced from
point-point spin correlators are coinciding with those from zero
momentum correlators. We investigate the contributions from
eigenvalues of the transfer matrix beyond the mass gap and discuss
the limitations and possibilities of such an analysis. The finite
size behaviour of the obtained $2d$ Ising model matrix elements
is examined. The point-to-point correlator formula is then applied
to Polyakov loop data in finite temperature $SU(2)$ gauge theory. The
leading matrix element shows all expected scaling properties.  
Just above the critical point we find a Debye
screening mass $~\mu_D/T\approx4~$, independent of the volume.

\end{abstract}

\newpage

\vspace{0.5cm}
%%%%%%%%%%%%%%%%%%%%%%%%%%%%%%%%%%%%%%%%%%%%%%%%%%%%%%%%%%%%%%%%%%%

\section{Introduction}

%%%%%%%%%%%%%%%%%%%%%%%%%%%%%%%%%%%%%%%%%%%%%%%%%%%%%%%%%%%%%%%%%%%

The determination of the correlation length $\xi$ and the screening
mass $\mu_D$ from point-to-point correlation functions of the
Polyakov loop is a non-trivial task, especially close to the critical
point of lattice gauge theories. The difficulties are resulting on
one hand from finite volume effects due to the nearby transition
and on the other hand from the unknown parametrization of the heavy
quark potential in the non-perturbative regime.

 In the transfer matrix (TM) formalism the levels of the transfer matrix
provide an access to both $\xi$ and $\mu_D$ without the introduction
of an ansatz for the quark potential. The formalism was
first applied with success to the exact solution
of the $2d$ Ising model \cite{o,sml}. It proved to be very efficient
as well in the analysis of the $4d$ Ising model \cite{desy1,desy2} and
also in the investigation of zero momentum correlators in 
Yang Mills theories \cite{bb1,bb2}. The zero momentum ( or plane--plane )
correlation functions are very convenient quantities to evaluate,
since their TM form is simply exponential and the levels and
matrix elements of the transfer matrix may be obtained easily from fits. 
  
To make the advantages of the TM formalism also available for the
analysis of point-to-point correlation functions on $d-$dimensional spatial
lattices we derive in this paper the corresponding TM expressions.
In the $2d$ Ising model we test and confirm
then the validity of our TM formula by comparison to the
results obtained from plane-plane correlators. Simultaneously we are
able to determine, where - as a function of the coupling constant -
levels beyond the mass gap are of importance
and what can be expected from such an analysis.
 
We apply then the TM technique also to the case of the (3+1) 
dimensional $SU(2)$ gauge theory. Our special point of interest is
here, in contrast to the intention of an earlier paper \cite{em}
on the subject, the study
of the higher levels, in particular their connection to the screening
mass $\mu_D$, and their influence on the determination of the mass gap. 

 Let us consider $d-$dimensional spatial lattices with periodic
boundary conditions of size $N^{d-1}L$,
where $N$ denotes the number of points in each transverse direction
and $L$ that in one selected direction ( the $z-$direction ).
The lattice spacing $a$ is set to unity in the following. The partition
funtion is then

\eq
Z \equiv \mbox{Tr} \left( {\bf V}^{L} \right),
\en

\noi and $~{\bf V}~$ is the transfer matrix in $~z-$ direction.
Its eigenstates $~\mid \! n \rangle~$ are chosen to be orthonormal.
We order them such that we have for the
eigenvalues $~\lambda_{n}~(n=0,~1,~2,~...)~$

\eq
{\bf V} \mid \! n \rangle = \lambda_{n} \cdot \mid \! n \rangle~;
\quad \lambda_{n} \equiv e^{-\mu_{n}} ;
                                                    \label{lambd0}
\en

\eq
\mu_{0} < \mu_{1} < \mu_{2} < ...~.
\en
 In addition we normalize our partition
function so that we have for the vacuum state

\eq
\lambda_0 = 1,~\mu_0 = 0 .
\en
This implies

\eqa
Z & = &
\sum_{n} \langle n \! \mid  {\bf V}^{L} \mid \! n \rangle
\nonumber \\ \nonumber \\
  & = & 1 + e^{-\mu_{1} L} + e^{-\mu_{2} L}
    + e^{-\mu_{3} L} + \ldots~,
                                                 \label{part1}
\ena
and the partition function varies then between 1 and 2.
\vspace{0.5cm}

%%%%%%%%%%%%%%%%%%%%%%%%%%%%%%%%%%%%%%%%%%%%%%%%%%%%%%%%%%%%%%%%%%%

\section{Plane--plane and point--point correlators}

%%%%%%%%%%%%%%%%%%%%%%%%%%%%%%%%%%%%%%%%%%%%%%%%%%%%%%%%%%%%%%%%%%%

\subsection{Plane--plane correlators}

%%%%%%%%%%%%%%%%%%%%%%%%%%%%%%%%%%%%%%%%%%%%%%%%%%%%%%%%%%%%%%%%%%%
We define zero momentum operators by

\eq
 \tilde {{\cal O}}(z) = N^{-\frac{d-1}{2}} \! \cdot
 \sum_{\vec{x}_{\perp}} {\cal O}(\vec{x}_{\perp},z) ~,
                                                \label{otilde_is}
\en

\noi where
$~{\cal O}(\vec{x}_{\perp},z)~$ is the Polyakov loop
$~{\cal P}(\vec{x}_{\perp},z)~$ for the $3+1$ dimensional $~SU(2)~$
gauge theory ($d=3$) and the spin $~\sigma_{x,z}~$ for the two
dimensional Ising model ($d=2$).
The corresponding correlation functions are

\eqa
\tilde{\Gamma} (z) & = &
\langle \tilde {{\cal O}}(z) \cdot \tilde { {\cal O}}(0) \rangle
 ~\equiv~  Z^{-1} \cdot Tr \left[ \tilde {{\cal O}}(0) \cdot {\bf V}^{z}
\cdot \tilde {{\cal O}}(0) \cdot {\bf V}^{L-z} \right]
\nonumber \\ \nonumber \\
& = & Z^{-1} \cdot \sum_{n < m} c_{mn}^2
\cdot \Bigl[ e^{- \mu_{m} z} \cdot e^{- \mu_{n} (L-z)}
+ e^{- \mu_{n} z} \cdot e^{- \mu_{m} (L-z)} \Bigr]~,
\nonumber \\ \nonumber \\
& = & Z^{-1} \cdot
\sum_{n < m} c_{mn}^2  \cdot e^{-\mu_n L}
\Bigl[ e^{- \mu_{mn} z}
+ e^{- \mu_{mn} (L-z)} \Bigr],
                                                 \label{cor1_is}
\ena

\noi where
\eq
\mu_{mn} \equiv \mu_m - \mu_n ;
~~c_{mn} \equiv
\langle n \! \mid \tilde{{\cal O}}(0) \mid \! m \rangle~,
\en

\noi are the level differences and the transition matrix elements.
Because the eigenstates of the transfer matrix are either
symmetric or antisymmetric under
tranformations, which change the sign of ${\cal O}$,
we have $~c_{nn} = 0~$.

Let us compare the contributions of the different states to the
correlator $~\tilde{\Gamma} (z)~$ above and
below the critical point $~\beta_{c}~$, where $~\beta~$ is
the coupling (to be specified below).
To do this we show explicitly the contributions to $~\tilde{\Gamma}(z)~$
associated with the three lowest nonzero states
\newpage

\eqa
\tilde{\Gamma}(z) = Z^{-1} \cdot \Bigl\{
&   &
c^2_{10} \cdot \left[ e^{- \mu_1 z} + e^{- \mu_1 (L-z)} \right]
\nonumber \\ \nonumber \\
& + &
c^2_{30} \cdot \left[ e^{- \mu_3 z} + e^{- \mu_3 (L-z)} \right]
\nonumber \\ \nonumber \\
& + &
c^2_{21} \cdot e^{-\mu_1 L} \left[ e^{- \mu_{21} z}
+ e^{- \mu_{21} (L-z)} \right] + \ldots~~ \Bigr\}~~.
                                                  \label{cor2_is}
\ena

%%%%%%%%%%%%%%%%%%%%%%%%%%%%%%%%%%%%%%%%%%%%%%%%%%%%%%%%%%%%%%%%%%%

\subsection{Point--point correlators}

%%%%%%%%%%%%%%%%%%%%%%%%%%%%%%%%%%%%%%%%%%%%%%%%%%%%%%%%%%%%%%%%%%%

In our subsequent analysis we establish a connection between the 
plane--plane and point--point correlators
\eq
\Gamma (\vec{x}) = 
\langle {\cal O}(\vec{x}) \cdot {\cal O}(0) \rangle~,   
                                           \label{popoc}
\en
in the context of the transfer matrix formalism.

The Fourier transform of the plane--plane correlator

\eq
\tilde{\Gamma} (p_{z}) = \sum_{z} e^{izp_z} \cdot \tilde{\Gamma} (z),
\en
is, using eq.(\ref{cor1_is})
\eq
\tilde{\Gamma} (p_{z}) = Z^{-1} \cdot \sum_{n < m} c_{mn}^{2} \cdot
e^{-\mu_{n} L} \cdot
 \tilde{G}(p_{z}; \mu_{mn})~,
\en

\noi where

\eq
\tilde{G}(p_{z}; \mu) \equiv
2 \left(1-e^{-\mu L} \right) \sinh \mu \cdot
\left[ 4 \sinh^{2} \frac{\mu}{2} +
4 \sin^{2} \frac{p_{z}}{2} \right]^{-1}~,
\en

\noi is just the Fourier transform of the sum of exponentials
$~[e^{-\mu z} + e^{-\mu (L-z)}]~$.
The connection between $~\tilde{\Gamma}(z)~$ and the point--point
correlator $~\Gamma(\vec{x}_{\perp};z) \equiv \Gamma(\vec{x})~$ is

\eq
\tilde{\Gamma} (z) =  \sum_{\vec{x}_{\perp}} \Gamma(\vec{x}_{\perp};z)~,
\en

\noi and their Fourier transforms are related by

\eq
\tilde{\Gamma}(p_{z}) \equiv \Gamma(\vec{p}_{\perp}=0,p_{z})~.
                                                         \label{cor3_is}
\en

\noi To obtain the full correlator
$~\Gamma(\vec{p}_{\perp},p_{z}) \equiv \Gamma(\vec{p})~$ we
use the substitution

\eq
4 \sin^{2} \frac{p_{z}}{2} \longrightarrow
D(\vec{p}) \equiv \sum_{i=1}^{d} 4 \sin^{2} \frac{p_{i}}{2}~,
                                           \label{ansatz}
\en

\noi where $~D(\vec{p})~$ is the lattice laplacian in
$~d$--dimensional momentum space. Therefore we arrive at

\eq
\Gamma (\vec{p}) = Z^{-1} \cdot \sum_{n < m} c_{mn}^{2} \cdot
e^{-\mu_{n} L} \cdot
 G(\vec{p}; \mu_{mn})~,
\en

\noi where

\eq
G(\vec{p}; \mu) =
2 \left(1-e^{-\mu L}\right) \sinh \mu \cdot
\left[ 4 \sinh^{2} \frac{\mu}{2} + D(\vec{p}) \right]^{-1}~,
\en
\noi and
\eq
\tilde {G}(p_z; \mu) = G(\vec{p}_{\perp}=0,p_z; \mu )~.
\en

\noi Finally, to get the correlator $~\Gamma (\vec{x})~$ we
perform the inverse Fourier transformation,
resulting in

\eqa
\Gamma (\vec{x}) =
Z^{-1} \cdot \Bigl\{ &   & c^2_{10} \cdot G(\vec{x};\mu_1)
\nonumber \\ \nonumber \\
                     & + & c_{30}^{2} \cdot G(\vec{x};\mu_3)
\nonumber \\ \nonumber \\
                     & + & c_{21}^{2} \cdot e^{- \mu_1 L} \cdot
                           G(\vec{x};\mu_{21}) + \ldots~ \Bigr\}~,
                                                         \label{cor4_is}
\ena

\noi and $~G(\vec{x};\mu)~$ is just the Fourier transform of
$~G(\vec{p}; \mu)~$

\eq
G(\vec{x}; \mu) =
\frac{ 1 }{N^{d-1}L} \cdot \sum_{\vec{p}} e^ {-i\vec{x}\cdot\vec{p}}
 \cdot G(\vec{p}; \mu)~.
                                                       \label{gear}
\en

The ansatz we used in  eq.(\ref{ansatz}) to obtain the
point--point correlator is so far without proof. Yet the results of
our numerical analysis of correlators strongly
support it (see below).

An expression for the expectation value of the square of the
lattice average of the operator
${\cal {O}} (\vec{x}) $ in terms of the matrix elements $c_{mn}$ may be
easily derived in the following way. Since

\eq
N^{d-1}L \langle {\cal {O}}^{2} \rangle~=  \sum_{\vec{x}} \Gamma(\vec{x})
 = \Gamma(\vec{p}=0),
\en
we obtain
\eq
N^{d-1}L \langle {\cal {O}}^{2} \rangle~= Z^{-1} \cdot \sum_{n < m}
 c_{mn}^{2} \cdot e^{-\mu_{n} L} \cdot
 G(\vec{p}=0; \mu_{mn})~,
                                                       \label{msqr}
\en

\noi with

\eq
 G(\vec{p}=0; \mu) =
 \left(1-e^{-\mu L}\right) \coth \frac{\mu}{2} ~.
                                                       \label{gsqr}
\en

The correlator $~\Gamma(\vec{x})~$ can be represented in the form
of a superposition of two Yukawa-type potentials only in the
case of the large-volume limit. If $~\mu \! \cdot \! N \sim 1~$
then the
finite-volume corrections are too strong and such a representation
is not possible.

%%%%%%%%%%%%%%%%%%%%%%%%%%%%%%%%%%%%%%%%%%%%%%%%%%%%%%%%%%%%%%%%%%%

\subsection{Correlation length and scaling behaviour}

%%%%%%%%%%%%%%%%%%%%%%%%%%%%%%%%%%%%%%%%%%%%%%%%%%%%%%%%%%%%%%%%%%%

Below the critical point $~\beta < \beta_{c}~$ the lowest nonzero
energy level $~\mu_1~$ - the mass gap - determines
the large distance behaviour of the correlation functions.
We therefore define the correlation length at
$~\beta \stackrel{<}{\sim} \beta_{c}~$ by (see also ref.\cite{barb})

\eq
\xi_{-}(\beta) \equiv \mu_1^{-1}
\sim \mid \! \beta - \beta_{c} \! \mid ^{-\nu}.
                                                 \label{xi}
\en

\noi The contribution of the next level with energy
$~\mu_2~$ is assumed to be suppressed
because of the additional factor $~e^{- \mu_1 L}~$
(or $~e^{- \mu_2 L} ~$)
in the third term of the right hand sides of eqs.(\ref{cor2_is})
and (\ref{cor4_is}). Therefore the third
level with energy $~\mu_3~$ gives the next to
leading corrections at large distances.

The situation is different at $~\beta > \beta_{c}~$ . There the mass
gap $\mu_1 \approx 0$ , if $~N~$ is large enough
and $~L \sim N~$ . In this case the first term on the right hand sides
of eqs.(\ref{cor2_is}) and (\ref{cor4_is}) becomes $~z-~$independent
and the large
distance behaviour is given by the next level difference
$\Delta\mu = \mu_{21} \approx \mu_2$ or $\mu_{30} = \mu_3$ ,
 so that the Debye mass is
\eq
\mu_D \equiv \Delta\mu~.
\en
Here $\mu_D$ is the nonperturbative equivalent to $2m_D$, 
where $m_D$ is the perturbative screening mass.
Near the phase transition point $~\beta \sim \beta_{c}~$
all three levels are expected to give an essential contribution.

In the thermodynamic limit below the phase transition point,
$~\beta < \beta_{c}~$, the correlator $~\Gamma({\vec x})~$ decays
exponentially at large distances $\mid \! \vec{x} \! \mid  \gg 1 $

\eq
\Gamma ({\vec x}) \sim \exp (-\mid \! \vec{x} \! \mid/\xi_{-}(\beta))~,
\en

\noi which entails

\eq
\sum_{\vec{x}_{\perp}} \Gamma(\vec{x}_{\perp};z) < \infty .
                                              \label{asymp1}
\en

\noi From eq.(\ref{asymp1}) we conclude that for a finite
lattice size all matrix elements are independent on $~N~$
in the large volume limit so that

\eq
      c_{mn}^{2} \sim N^{0}~; ~~~~~~N \rightarrow \infty .
                                              \label{casy1}
\en

Well above the transition point, $~\beta > \beta_{c}~$, the
behaviour of the correlator $~\Gamma({\vec x})~$
for large separations $\mid \! \vec{x} \! \mid  \gg 1$ is

\eq
\Gamma ({\vec x}) - a \sim
\exp (-\mid \! \vec{x} \! \mid/\xi_{+}(\beta))~;
 ~~~~N \rightarrow \infty,
\en

\noi where $a$ is a positive constant, independent of
 $\mid \! \vec{x} \! \mid $ and
of $~N~$ for large but finite $~N~$ so that

\eq
\sum_{\vec{x}_{\perp}} \Gamma(\vec{x}_{\perp};z) =
\tilde{\Gamma} (z) \sim N^{d-1}.
                                                  \label{asymp2}
\en
 
\noi The constant is connected to the existence of a nonvanishing
spontaneous magnetization and is in fact equal to
$~\langle {\cal {O}}^{2} \rangle~$ on finite lattices \cite{grif} .
\noi As the major contribution to $~\tilde{\Gamma} (z)~$ is coming
from the first term in eq.(\ref{cor2_is}), which is proportional
to $~c^2_{10}~$ and essentially independent of $~z~$, since
$~\mu_1L \ll 1~$. We expect therefore that

\eq
 c_{10}^{2} \sim N^{d-1}; ~~~~~~N \rightarrow \infty .
                                              \label{casy2}
\en

\noi The same conclusion may be drawn as well directly by assuming 
$~N-$independence of $~\langle {\cal {O}}^{2} \rangle~$ and
eqs. (\ref{msqr}) and (\ref{gsqr}).

In the very neighbourhood $\beta \sim \beta_c$ of the phase
transition we may apply
finite size scaling techniques \cite{barb,fish} to derive the
$~N-$dependence of the matrix elements. Let us assume for
simplicity, that $~N=L~$.
According to finite size scaling theory any observable $O$ with critical
behaviour is supposed to have the following form

\eq
O = N^{\rho /\nu} \cdot f_O (x N^{1/\nu})~;
~~~~N \rightarrow \infty,
                                              \label{fss0}
\en

\noi for fixed small $~x \equiv (\beta - \beta_{c})/\beta_{c}~$. Here
$\rho$ is the critical exponent of the observable $O$ and $\nu$ the one
of the correlation length. Due to eq.(\ref{xi}), we expect then that
the mass gap behaves for $~\beta \leq \beta_{c}~$ ($x \leq 0$) as

\eq
\mu_1 = N^{-1} \cdot
f_{\mu}(x N^{1/{\nu}})~;
~~~~~x \sim 0 ~.                               \label{fss1}
\en

\noi In the same $\beta-$region the susceptibility
$~\chi~$ is defined as follows

\eq
 \chi = N^{d} \langle {\cal O}^{2} \rangle~;
~~~~\beta \leq \beta_{c} ~.
                                            \label{chi}
\en

\noi Its critical behaviour is governd by the exponent $\gamma$ so
that

\eq
N^d \langle {\cal O}^{2} \rangle = N^{\gamma /\nu}
\cdot f_{\chi}(x N^{1/\nu})~;
~~~~x \sim 0 ~.
                                             \label{fss4}
\en

\noi Since again the leading contribution to
$~\langle {\cal O}^{2} \rangle~$ is proportional to the matrix
element $~c_{10}^2~$ we find after combining the last equation with
eq.'s (\ref{msqr}), (\ref{gsqr}) and (\ref{fss1}) that

\eq
c_{10}^{2} = N^{-1 +\gamma /\nu} \cdot
f_{c}(x N^{1/{\nu}})~; ~~~~~x \sim 0 ~.
                                              \label{fss6}
\en

\noi Although the latter equation was formally derived only
for $\beta \leq \beta_c$, we expect, because of the analytic
$\beta-$dependence on finite lattices,
that it will be valid also above the critical point.

%%%%%%%%%%%%%%%%%%%%%%%%%%%%%%%%%%%%%%%%%%%%%%%%%%%%%%%%%%%%%%%%%%%

\section{Numerical results}

%%%%%%%%%%%%%%%%%%%%%%%%%%%%%%%%%%%%%%%%%%%%%%%%%%%%%%%%%%%%%%%%%%%

\subsection{The twodimensional Ising model}

%%%%%%%%%%%%%%%%%%%%%%%%%%%%%%%%%%%%%%%%%%%%%%%%%%%%%%%%%%%%%%%%%%%

\noi The $2d$ Ising model provides an ideal test case for a
comparison of point-point and plane-plane correlators. This is so,
because the levels $\mu_n$ are all explicitly known \cite{o,sml}.
Therefore a fit of the correlators in terms of the TM formulae,
eqs. (\ref{cor2_is}) and (\ref{cor4_is}), requires only the determination
of the matrix elements. Moreover, Monte Carlo simulations of the
model are relatively simple.

To be more specific, consider a $~d=2~$ Ising system on a lattice
of size $~L_{x} \cdot L_{z} \equiv N \cdot L~$ with periodic boundary
conditions. At every site $~i \equiv (x,z)~$ there is a spin
$~\sigma_{i} = \pm 1~$. The partition function $~Z~$ is of the form

\eq
Z = \sum_{\{ \sigma_{k}=\pm 1 \} }
\exp \left( \beta \cdot \sum_{<ij>} \sigma_{i}\sigma_{j} \right),
\en

\noi where $~\beta~$ is the inverse temperature and $~<ij>~$ means
that only nearest neighbours interact .

The eigenvectors $~\mid \! n \rangle~$ of the transfer matrix
as well as the eigenvalues
$~\lambda_{n}~$ correspond to different numbers of collective
excitations (quasiparticles) in transverse direction with
momenta $~q_{1},q_{2}, \ldots~$. These momenta take values
(for even $~N$)
\eqa
q & = &  \pm \frac{\pi}{N}, ~\pm \frac{3\pi}{N},~\ldots,
~\pm \frac{\pi (N-1)}{N} ~
\quad \mbox{for an even number of quasiparticles}
\nonumber \\
q & = &  0,~\pm \frac{2\pi}{N}, ~\ldots,
~\pm \frac{\pi (N-2)}{N} ,~\pi~
\quad \mbox{for an odd number of quasiparticles.}
\nonumber
%                                                   \label{mom}
\ena
\noi We may characterize the eigenvectors $~\mid \! n \rangle~$
via the momenta of the quasiparticles
$~q_{1}<q_{2} < \ldots <q_{m}~$ as
$~\mid \! k_{1},k_{2}, \ldots, k_{m} \rangle~$ where
$~q \equiv \frac{2\pi}{N} k~$.
Of course, the eigenvectors and eigenvalues depend on $~N~$ but
do not depend on $~L~$.

The four lowest states are then the vacuum
$~\mid \! 0 \rangle~$, $~\mid \! 1 \rangle~=~\mid \! k_{1}=0 \rangle~$,
$~\mid \! 2 \rangle~=~\mid \! -\frac{1}{2};\frac{1}{2} \rangle~$
and $~\mid \! 3 \rangle~=~\mid \! -1;0;1 \rangle~$ .
Above the vacuum level $~\mu_0 \equiv 0~$,
the next smallest $~\mu~$-- the mass gap $~\mu_1~$-- is
nonzero in the limit $~N \ra \infty~$
below the critical point $~\beta < \beta_{c}~$.
At the critical point $~\mu_1(\beta_{c}) \sim N^{-1}~$,
as expected from eq.(\ref{fss1}), and
above the critical point $~\mu_1~$ tends to zero with
increasing $~N~$ as $~\mu_1 \sim N^{-1/2} e^{-\kappa \cdot N}~$.
In the thermodynamic limit the vacuum becomes therefore
degenerate. In Fig. 1 we show for $N=30$
the dependence on $~\beta/\beta_c~$
of the levels $\mu_1$ to $\mu_5$ and the smallest and therefore
most relevant level differences, which appear in the correlator
formulae.

To test the substitution, eq.(\ref{ansatz}), and the resulting eq.(
\ref{cor4_is}) for the correlator $~\Gamma(\vec{x})~$,
we have measured plane--plane and point--point correlators
of the spin operator $~\sigma_i~$
on $N=L=30,40,50,60$ lattices. At each point $500000$ cluster updates
were performed and measurements taken every 10th update.
Subsequently we have carried out fits to both correlators with
varying numbers of levels to obtain the matrix elements.
The results for the matrix elements on an $~N=L=30~$ lattice are
compared in Fig. 2. In each case we show fits including either
all distances $~r=\mid \! \vec{x} \! \mid=1,2,...,15~$ or only
those for $~r>2~$. If more levels than assumed in the fit are
contributing, we see a dependence on the lowest distance $r$
taken into account in the fit. The effect is more pronounced
for the point--point correlator than for the plane--plane correlator.
Summarizing the experiences we have made with the different fits,
we observe, that both formulae lead to exactly the same
results, whenever the maximal number of levels is taken into
account, which lead to non-negative $~c_{mn}^2~$, i.e. our ansatz
is definitely confirmed.

The final result for the $2d$ Ising model and $~N=L=30~$ is shown
in Fig. 3. We find that for $\beta < 0.92\beta_c$ only one term with
 $\mu_{10}=\mu_1$, the mass gap, contributes; near $\beta_c$ up to
three terms are essential and well above the critical point,
for $~\beta>1.1\beta_c~$, where the mass gap
 $\mu_1 \approx 0$ only one more term is present. The matrix
 element $~c_{21}^2~$ is increasing below $\beta_c$ with
 decreasing $\beta$. However, the relevant factor in the correlator
 formulae, $~c_{21}^2\exp(-\mu_1 L)~$, is negligible below
 $~\beta = 0.92\beta_c~$, so that $~c_{21}^2~$ can no longer be
 determined from the fits.

We have also studied the scaling properties of the major matrix
element $~c_{10}^2~$ in the three different $\beta-$regions.
As can be seen from Fig. 4, the predictions, eqs.(\ref{casy1}),
(\ref{casy2}) and (\ref{fss6}) are all nicely confirmed by our results.
Here we have used the known $2d$ Ising model values for $\gamma=7/4$
and $\nu=1$.
Obviously, we could not check the $~N-$independence
according to eq.(\ref{casy1}) of the higher matrix elements such
as $~c_{21}^2~$, since - as mentioned - their contributions are
negligible for $\beta$ well below $\beta_c$. 

%%%%%%%%%%%%%%%%%%%%%%%%%%%%%%%%%%%%%%%%%%%%%%%%%%%%%%%%%%%%%%%%%%%

\subsection{SU(2) lattice gauge theory}

%%%%%%%%%%%%%%%%%%%%%%%%%%%%%%%%%%%%%%%%%%%%%%%%%%%%%%%%%%%%%%%%%%%

We now want to apply our TM formula for the point--point correlator
to $SU(2)$ gauge theory. We consider a finite lattice of size
$~L_{t} \cdot L_{\perp}^{d-1} \cdot L_{z} \equiv
 L_{t} \cdot N^{d-1} \cdot L~$
($d=3$) with periodic boundary conditions.
The standard Wilson action for $~SU(2)~$ gauge theory is

\eq
 S_{W} = \beta \cdot \sum_{\Box} \left( 1 -
\frac{1}{2} Tr U_{\Box}\right),
\en

\noi where $~\beta = 4/g^{2}~$ and $~\Box \equiv (x; \kappa \rho )~$
refers to location and orientation of the plaquette. The field variables 
$~U_{\kappa}(\vec{x};t) \in SU(2)~$ are
defined on the links, and the $~U_{\Box}~$ are plaquette variables
$~
U_{\Box} \equiv U_{x; \kappa \rho} =
 U_{x; \kappa }U_{x+\kappa; \rho }U_{x+\rho; \kappa }^{\dag}U_{x; \rho }^{\dag}
~$.
For the Wilson action the transfer matrix $~{\bf V}~$ is
proven to be positive definite \cite{crl,lu1}. Also,
due to the Perron-Frobenius theorem \cite{fp} the vacuum
$~\mid \! 0 \rangle~$ is unique for a finite system, and we can
choose again the normalization such that $~\mu_{0} \equiv 0~$ .

Here, the Polyakov loop $~{\cal P}(\vec{x})~$ takes the r\^ole
of the operator $~{\cal O}(\vec{x})~$. It
is defined as usual by

\eq
\quad
 {\cal P}(\vec{x}) \equiv \frac {1}{2} \mbox{Tr} \left[
\prod_{t =1}^{L_{t}} U_{4}(\vec{x},t) \right].
\en

The Monte Carlo data for the point--point correlator,
which we want to analyze in the following,
were computed \cite{eng} on $~L_t=4,N=L=12,18,26~$ lattices
with $~10^5-4 \cdot 10^5~$ updates and measurements every 10th
sweep. In contrast to the case of the $2d$ Ising model, in $SU(2)$
gauge theory the level differences are unknown and have, like the
matrix elements, to be determined through the fit.

In general we find a behaviour resembling very much the one of the 2$d$
Ising model. In particular, the number of levels, which may be extracted
from the fits is comparable. Fits with more than two levels are only
possible on the largest lattice very close to the transition. Otherwise
one either obtains negative squares of matrix elements or there is no
minimum of $\chi^2$. Taking into account more than one term in eq.
(\ref{cor4_is}) tends to decrease the result for the mass gap level $\mu_1$.
This can be seen in Fig. 5, where we show the results for $N\mu_1$ from
one and two level fits. The inclusion of a third level in the fit, however,
does not change $\mu_1$ anymore.
  
The fit result $c_{10}^2/Z$ for the major matrix element was subsequently 
checked for its scaling properties. Note, that here the partition function
$Z$ is not explicitly calculable from eq. (\ref{part1}), because we know 
only the lowest level(s). Like in the case of the $2d$ Ising model we find 
all predictions from eqs. (\ref{casy1}), (\ref{casy2}) and (\ref{fss6})
very well confirmed. This is shown in Fig. 6. In the scaling test, Fig. 6b,
we have used $\gamma=1.24$ and $\nu=0.63$, the values of the $3d$ Ising
model, in accord with the universality hypothesis \cite{sy}, which predicts
equal critical exponents for $SU(2)$ and the $3d$ Ising model. 

It is interesting to look at the behaviour of the next to leading
level ( or level difference ) $\Delta\mu$.
 As can be seen from Fig. 7, $\Delta\mu$ drops from a higher value
below $\beta_c$ at the transition to a value near to one ( in lattice
units ) and stays then relatively constant and moreover independent
of the lattice sizes used here. This second level fixes the large
distance behaviour 
of the correlation functions above $\beta_c$, since $\mu_1$, as
is evident from Fig. 5, is essentially zero there and a third level
does not contribute outside the transition region.
 Therefore we identify it with $\mu_D$.
Because we have $L_t=4$ and $T=1/L_t$ we are led to a ratio $\mu_D/T\approx 4$,
slightly higher than the ratio found with conventional
methods\cite{debye4}. On the other hand a higher value seems to be
preferred by next-to-leading order perturbation theory calculations
\cite{debye5}. In the close vicinity of the transition we have found
at four $\beta-$values solutions to three level fits on the $N=26$
lattice. They are also shown in Fig. 7. We see that the lower of the
two levels beyond the mass gap is approaching zero at the transition.
Indeed, if interpreted as $\mu_D$, the expected $N-$behaviour at the 
transition is propotional to $~N^{-1}$.
  
Finally we present in Fig. 8 the second matrix element $c^2_2/Z$
resulting from two level fits. The fluctuations of $c^2_2/Z$ in the
neighbourhood of the transition are probably due to both the
statistical uncertainties in the data and the possible influence
of higher levels. It is remarkable, that outside the transition region 
the matrix element is $N-$independent, constant and moreover about
equal well above and below the critical point. Compared to the first 
matrix element, however, the second one is rather small.

%%%%%%%%%%%%%%%%%%%%%%%%%%%%%%%%%%%%%%%%%%%%%%%%%%%%%%%%%%%%%%%%%%%

\section{Summary}

%%%%%%%%%%%%%%%%%%%%%%%%%%%%%%%%%%%%%%%%%%%%%%%%%%%%%%%%%%%%%%%%%%%

We have derived a transfer matrix formula for the point-to-point
correlation function on $d-$dimensional spatial lattices.
The advantage of such an approach lies in its direct access
to the correlation length and/or the screening mass. Moreover,
the disconnected point-point correlation functions may be analysed
without the need for any subtraction.
 
The formula was tested in the $2d$ Ising model by comparison with
plane-plane correlation functions. We find that from both observables
the same information may be extracted, whenever the maximal number
of levels are taken into account, which lead to physically 
meaningful results in the fitting procedure. Quite naturally a
difference is observed, if existing higher level contributions are
neglected.

In both the $2d$ Ising model and the $(3+1)$ dimensional $SU(2)$
gauge theory we found the same general behaviour of the levels and
matrix elements :

\begin{enumerate}
\item well below the critical point only the mass gap $\mu_1$
is contributing to the correlators, the matrix element $c_{10}^2$
is independent of $N$ ;
 
\item close to the transition up to three levels are contributing
and $c_{10}^2 \sim N^{-1+\gamma/\nu}$ ;
 
\item well above the critical point only one higher level beyond
the essentially zero mass gap is contributing and $c_{10}^2 
\sim N^{d-1}$ .
\end{enumerate}
 
The detection of still higher levels seems to require an extremely
large statistics of the data and very large lattice volumes.
Most probably there is only a chance for such a program close to
the transition point.
 
In $SU(2)$ gauge theory we have calculated the change in the mass
gap due to the presence of the higher levels. We find that this
effect decreases the mass gap result. Finally we have determined the 
Debye screening mass to $\mu_D/T \approx 4$, independent of the 
lattice size used.

\newpage

{\small

%%%%%%%%%%%%%%%%%%%%%%%%%%%%%%%%%%%%%%%%%%%%%%%%%%%%%%%%%%%%%%%%%%%

\vfill
\eject
   
\newpage
 
%%%%%%%%%%%%%%%%%%%%%%%%%%%%%%%%%%%%%%%%%%%%%%%%%%%%%%%%%%%%%%%%%%%
 
{\large {\bf  Figure captions.}}

%%%%%%%%%%%%%%%%%%%%%%%%%%%%%%%%%%%%%%%%%%%%%%%%%%%%%%%%%%%%%%%%%%%
\vspace{.5cm}
\noi {\bf Fig.1} The lowest energy levels in the $2d$ Ising model
as a function of $~\beta/\beta_c$ for $~N=30$. The mass gap
$~\mu_1$ and $~\mu_2$ up to $~\mu_5$ are shown as solid lines,
the differences $~\mu_{21}=\mu_{2}-\mu_{1}$
and  $~\mu_{32}=\mu_3-\mu_2~$ as dashed lines.

\vspace{.5cm}
\noi {\bf Fig.2} Comparison of matrix element fits with one level
({\bf a}), two levels ({\bf b}) and three levels ({\bf c}) in the
$N=L=30~~2d$ Ising model vs. $~\beta/\beta_c$.
 The results shown are from fits
to zero momentum correlators including all distances for $~r>0$
(dashed lines) and $~r>2$ (solid lines). The corresponding results
for point-to-point correlators are shown by dotted and dashed-dotted
lines.

\vspace{.5cm}
\noi {\bf Fig.3} The lowest levels $~\mu_1,~\mu_3$ and $~
\mu_{21}$ (dashed lines) and the corresponding best fit matrix elements
$~c^2_{10},c_{30}^2~$ and $~c_{21}^2e^{-\mu_1 L}~$
as a function of $~\beta/\beta_c~$ in the $~N=L=30~~2d~$ Ising model.
The dotted line is $~c_{21}^2$.

\vspace{.5cm}
\noi {\bf Fig.4} Scaling properties of the mass gap matrix element
$~c^2_{10}~$ below $~\beta_c~$ ( $\sim N^0~$ ), close to $~\beta_c~$
( $\sim N^{\gamma/\nu-1}~$ as a function of $~xN~,
 x=(\beta-\beta_c)/\beta_c~$ )
and above $~\beta_c~$ \break ( $\sim N~$) for
$~N=L=30,40,50,60~$~(solid,dashed,dashed-dotted,dotted lines)~
in the $~2d~$ Ising model.

\vspace{.5cm}
\noi {\bf Fig.5} ~~The dependence of
$~N\mu_1~$ on the number of levels used in the fit for $~N=18~$
(squares) and $~N=26~$ (diamonds) as a function of $~\beta~$
in $~SU(2)~$ gauge theory. Two level fits are shown by filled symbols,
one level fits by empty ones. The inset shows for $~N=26~$
also three level fits (circles).

\vspace{.5cm}
\noi {\bf Fig.6} ~Scaling properties of the mass gap matrix element
$~c^2_{10}/Z~$ in $~SU(2)~$ gauge theory. The figure corresponds to Fig. 4.
Here $~N=12,18,26~$~(crosses,squares,diamonds).

\vspace{.5cm}
\noi {\bf Fig.7} ~The level difference $~\Delta\mu~$ from two level fits
in $~SU(2)~$ gauge theory as a function of $~\beta~$. The notation
is the same as in Fig. 6. The results of three level fits on the $N=26$
lattice for the two levels beyond $\mu_1$ are shown as filled diamonds.

\vspace{.5cm}
\noi {\bf Fig.8} ~The second matrix element $~c_2^2/Z~$ from two level
fits in $~SU(2)~$ gauge theory as a function of $~\beta~$. The notation
is the same as in Fig. 6.

\vfill
\eject

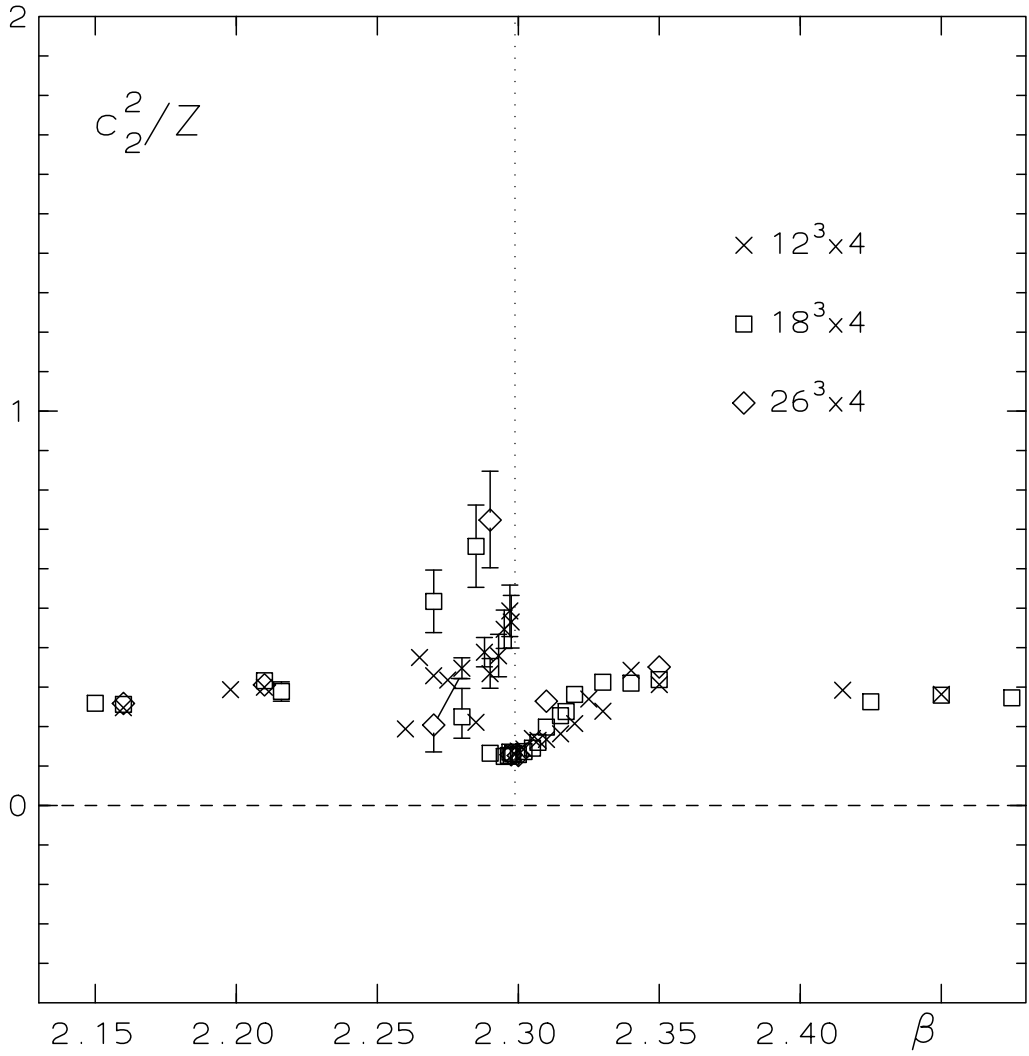
\end{document}